\documentclass[12pt,preprint]{aastex}

\shorttitle{Spectroscopy of Molecular Hydrogen From KH~15D}
\shortauthors{Deming et al.}

\begin{document}


\title{Spectroscopy of Molecular Hydrogen Emission From KH~15D\altaffilmark{1}}

\altaffiltext{1}{Data presented herein were obtained at the W.M. Keck Observatory, which is 
operated as a scientific partnership among the California Institute of Technology, 
the University of California and the National Aeronautics and Space
Administration. The Observatory was made possible by the generous financial 
support of the W.M. Keck Foundation.}

\author{Drake Deming}
\affil{NASA's Goddard Space Flight Center, \\
Planetary Systems Branch, Code 693, Greenbelt MD 20771}
\email{ddeming@pop600.gsfc.nasa.gov}

\and

\author{David Charbonneau}
\affil{California Institute of Technology, 105-24 (Astronomy),\\ 
1200 E. California Blvd., Pasadena, CA 91125}
\email{dc@caltech.edu}

\and

\author{Joseph Harrington}
\affil{Center for Radiophysics and Space Research, Cornell University \\
 326 Space Sciences Bldg., Ithaca, NY 14853-6801}
\email{jh@oobleck.astro.cornell.edu}


\begin{abstract}

We report infrared spectroscopy of the unusual eclipsing pre-main
sequence object KH~15D, obtained using NIRSPEC on Keck II.  During
eclipse, observations using low spectral resolution
($\lambda/\delta\lambda \sim 1000$) reveal the presence of prominent
molecular hydrogen emission in 5 lines near $2~\mu$m.  The relative
line strengths are consistent with thermal excitation at $T \sim 2800
\pm300K$.  Observations out of eclipse, at both low and high spectral
resolution ($\lambda/\delta\lambda \sim 2 \times 10^{4}$), show
reduced contrast with the stellar continuum. The change in contrast
for the strongest line, 1-0~S(1), is consistent with an approximately
constant emission line superposed on a variable stellar continuum.
Emission in the 1-0~S(1) line is observed to extend by $\gtrsim 4$
arc-sec both east and west of the stellar point spread function
($\gtrsim 3000$ AU).  Observed at high spectral resolution, the
velocity and intensity structure of the 1-0~S(1) profile are both
asymmetric.  East of the stellar PSF (by $1.1 - 2.3$ arc-sec) the
emission is blueshifted ($-63$ km sec$^{-1}$), and has significantly
greater intensity than the marginally redshifted component ($+2$ km
sec$^{-1}$, $\sim$ consistent with zero) which dominates west of the
stellar PSF. The spatial extent of the emission, and the excitation
temperature, suggest shock-excitation of ambient gas by a bipolar
outflow from the star and/or disk.  However, it is difficult to
account for the observed radial velocity unless the outflow
axis is inclined significantly to the plane of the sky.
\end{abstract}



\section{Introduction}

KH~15D is a weak-lined T Tauri star (K7V) in NGC 2264 ($d=760$ pc),
which shows deep (3.5 mag) periodic eclipses, each lasting for about
20 days, a large fraction of the 48.36 day period
\citep{her02,hh03}. The eclipses occur abruptly, indistinguishable
from obscuration by a `knife edge' \citep{her02}.  During eclipse, the
lack of reddening in the spectrum \citep{ham01}, and polarization
which varies only weakly with wavelength \citep{agol03}, imply that
the obscuring matter must be comprised of large particles. A consensus
view to date is that the eclipses are caused by structure in a
protoplanetary disk, fortuitously edge-on to our line of
sight. Candidate structures include a density wave \citep{her02}, a
vortex \citep{bv03}, or a warp in the disk induced by planet-disk
interactions \citep{agol03,winn03}.  The considerable recent interest
in KH~15D derives from the diagnostic potential inherent in the
eclipse geometry, and the serendipity of finding a protoplanetary
system whose structure suggests that planetary formation is actively
occurring.

Although the eclipses in KH~15D are evidently caused by large solid
particles, observations of the gaseous component may help to deduce
the nature of the system.  \citet{ham03} observed double-peaked atomic
hydrogen emission lines, indicative of a bipolar outflow.  In this
{\it Letter} we report the spectroscopic discovery of $2~\mu$m
molecular hydrogen lines in emission, also showing a double-peaked
structure indicative of outflow.  This H$_2$ emission has been
independently discovered by \citet{tok03}, by imaging in a narrow-band
filter. Molecular hydrogen emission is commonly observed from pre-main
sequence objects \citep{bach96}, where it can be excited by
fluorescence \citep{black87} or by shocks \citep{sb82}, and has been
well-studied in many cases.  H$_2$ observations should therefore help
to relate KH~15D to other pre-main sequence systems.

\section{Observations}

We observed KH~15D using the NIRSPEC spectrometer on the Keck II
telescope \citep{mclean98}.  Observations on 19 \& 26 August 2002 UT
used NIRSPEC's low spectral resolution mode, with slit widths of
$0.76$ and $0.57$ arc-sec respectively, giving
${\lambda}/{\delta}{\lambda} \sim 1100$ and $\sim 2000$.  KH~15D was
in eclipse on 19 August (phase $=0.17$, emergence was just beginning; 
zero phase is center of eclipse), and out of eclipse on 26 August
(phase $=0.32$). Noise from the sky background was higher than usual
on both nights because the object was observed at high airmass
($\sim 2$) in pre-dawn conditions, with a rapidly brightening
sky. (KH~15D observations were supplementary to our main program.)
Sky subtraction was facilitated by nodding the telescope to well
separated `a' and `b' positions along the slit.  The position angle of
the slit (measured positive eastward from north), and total
integration time used, were $75^{\circ}$ \& $600$ seconds, and
$134^{\circ}$ \& $400$ seconds, on 19 \& 26 August, respectively.

On 2 September 2002 UT, KH~15D was out of eclipse (phase$=0.46$). We
observed it using NIRSPEC's high-resolution cross-dispersed mode,
giving ${\lambda}/{\delta}{\lambda}~\sim~23,000$, with a $0.43$
arc-sec slit width (2 pixels), in position angle $66^{\circ}$. Total
integration time was $1600$ seconds.  These high resolution
observations were made under better sky conditions (airmass
$\sim~1.7$), and also used nodding for sky subtraction. For both the
low and high resolution spectra, we observed a neon emission lamp for
wavelength calibration and a continuum lamp for flat-fielding.

\section{Spectral Analysis}

\subsection{Low Resolution Spectra}

Sky and dark current subtraction was accomplished to first order by
subtracting the `b' frames from the `a' frames, after division by the
flat-field lamp frame.  A second-order sky correction was accomplished
by subtracting a residual sky spectrum from the average difference
frame.  The residual sky spectrum was determined on the average
difference frame by summing a region located spatially between the `a'
and `b' positions of the star on the slit.  The KH~15D spectrum was
obtained in a standard extraction, by summing both the `a' and `b'
spectra over a region of $\pm 2$ arc-sec along the slit, after
correction for the tilt of the slit.  The wavelength scale was
determined by fitting Gaussian profiles to 14 neon emission lines in
calibration spectra taken immediately following the KH~15D
spectra. The neon rest wavelengths (from the NIRSPEC manual) were
expressed as a quadratic function of the fitted line positions (in
pixels), with coefficients determined by least squares. The precision
of the wavelength calibration, estimated from the scatter in the
quadratic fits, is $\sim 0.0002~\mu$m (0.5 pixels).  Figure~1 shows
the low-resolution spectra in and out of eclipse, with the wavelength
scale transformed to the stellar frame.  Our velocity transformations
assume a heliocentric stellar radial velocity of $+10$ km~sec$^{-1}$,
intermediate between the two values cited by \citet{ham03}.  Both
Figure~1 spectra have been flat-fielded, and normalized to a unit
intensity near $2.1~\mu$m.  We did not observe a standard flux source,
but we were able to use the KH~15D stellar continuum to estimate an
absolute line intensity (see below).

The low resolution spectrum during eclipse shows 5 emission lines due
to H$_2$, as noted on the Figure. Even the weakest of these lines,
2-1~S(1), is real, which we judged by inspecting the average
difference frame prior to spectral extraction.  On the difference
frame, emission in the strongest line, 1-0~S(1), has a spatial
distribution along the slit which is peaked at the stellar position to
within $0.2$ arc-sec (1 pixel).  This bright emission can be traced beyond the
stellar point spread function (PSF) by at least $\pm 4$ arc-sec. There
is evidence for faint S(1) emission (intensity $\sim 0.2$ of the KH~15D peak) at
one other position along the slit, $\sim 16$ arc-sec distant from KH~15D.  

Intensities for the H$_2$ lines were determined by fitting Gaussian
profiles.  The widths of the fitted Gaussians were all constrained to
equal the best-fit width for the 1-0~S(1) line, since the lines are
not spectrally resolved.  Errors in the line intensities (areas under
the Gaussian fits) were estimated using a Monte-Carlo procedure: we
repeated the fits many times on `synthetic' lines having the same line
parameters and noise levels as in the real data.  All of the line
intensities and errors are normalized to unit intensity for the
1-0~S(1) line, and are tabulated in Table~1.  We used the very high
spectral resolution atlas of \citet{liv91}, convolved to the $\sim 25$
km sec$^{-1}$ widths of the lines (see below), and measured the
telluric absorption at the Doppler-shifted position of each line.  At
the observed geocentric Doppler shift and line width, none of the
KH~15D lines are coincident with significant telluric lines, hence any
telluric corrections are negligible, even at 2 airmasses.

The weakness of the 2-1~S(1) line is indicative of thermal excitation,
since models of fluorescent emission predict an intensity for this
line of $\sim 0.5$ relative to the 1-0~S(1) line \citep{black87}.  It
is clear from Figure~1 that the intensity of 2-1~S(1) is substantially
less than half of the 1-0~S(1) line. To determine the excitation
temperature, we computed ln$(I/gA)$ where $I$ is the Table~1 line
intensity, $g$ is the upper level statistical weight, and $A$ is the
Einstein A-value.  For optically thin emission, ln$(I/gA)$ is
proportional to the energy of the emitting level, with constant of
proportionality $= -1/T$.  We adopted the line parameters tabulated by
\citet{fernan97}, and calculated $T~\sim~2800\pm300K$ by
least squares.  The relative intensities expected from thermal excitation at
$T=2800K$ are tabulated in Table~1 for comparison.  The differences
between these values and the observed intensities are somewhat greater
than our estimated observational errors, but not so discrepant as to
contradict thermal excitation.

\subsection{High Resolution Spectra}

Figure~1 shows that the emission line contrast is greatly reduced out
of eclipse, similar to the effect seen in atomic emission lines by
\citet{ham03}. In the presence of shot noise from the stellar continuum, 
observation of the emission lines outside of eclipse is problematic at
low spectral resolution. But the contrast for unresolved emission lines is
proportional to spectral resolution, and we have been able to measure
the 1-0~S(1) line in high resolution spectra outside of eclipse.

Our high resolution spectra were extracted from an average difference
image in a manner similar to the low-resolution spectra. However, the
high-resolution spectra did not require a second order sky
subtraction, and we used night sky emission lines for wavelength
calibration.  A velocity (wavelength) precision of $1.8$ km sec$^{-1}$
(0.4 pixels) was achieved by using four night sky lines, two on each
side of the 1-0~S(1) line, with vacuum wavelengths from \citet{rouss}.
We adopted the precise vacuum wavelength ($2.1218356~\mu$m) of the
1-0~S(1) line from \cite{bbs82}. Similar to the low resolution
in-eclipse spectra, our high resolution spectra also show the emission
extending well beyond the stellar PSF by $\gtrsim 4$ arc-sec.  We
extracted three high-resolution spectra from the average difference
frame, and these spectra are shown in Figure~2.  The first spectrum
(upper portion of Figure~2) was extracted by summing along the slit
within $\pm 0.9$ arc-sec of the star.  Two other spectra were
extracted over the interval $1.1-2.3$ arc-sec from the star to the
east and west (lower spectra in Figure~2).

We examined the total intensity of the line seen in Figure~2 (top
spectrum, near the star), compared to the line intensity seen near the
star during eclipse.  Using the stellar continuum as a reference, and
adopting an eclipse depth of 3.1 mag at the epoch of the Figure~1
spectrum (phase$=0.17$; \citet{her02}), we compute the intensity
expected in Figure~2 (out of eclipse) by scaling the Figure~1 (in
eclipse) line intensity.  Since the emission is extended, we
allow for the different solid angles (slit widths times distance
summed along the slit). On this basis we expect to see a line
intensity about $25\%$ less than is actually observed in Figure~2.
This is certainly within the errors of the observations, and we
conclude that the 1-0~S(1) line intensity is approximately constant,
and is not eclipsed.  The variation in H$_2$ emission line contrast
can be accounted for (within the errors) solely by changes in the
stellar brightness.

The line profiles seen in Figure~2 show two components. Relative to
the reference frame of the star, a blueshifted component appears at
$-63$ km sec$^{-1}$, and a redshifted component at $+2$ km sec$^{-1}$.
Within the errors, the redshifted component is consistent with zero
velocity.  Although the total intensity in the emission is peaked at
the stellar position, the relative intensity of the blue- and
redshifted components changes abruptly there.  East of the star, the
blueshifted component dominates, whereas the redshifted component
prevails to the west. The components are relatively narrow, FWHM $\sim
25$ km sec$^{-1}$.

We have estimated the absolute line intensity for the Figure~2
spectrum coincident with the stellar position, adding both components.
Using the $V$ magnitude reported by \citet{ham01}, and the $V-K$ color
normal to a K7 dwarf from \citet{bessel}, we deduce a $K$ magnitude
for the star of 12.9, outside of eclipse. Adopting the zero-magnitude
flux from \citet{bessel}, we calculated the line intensity using the
stellar continuum as a photometric standard.  Since the star is a
point source, whereas the emission is extended, we corrected for
stellar slit-losses. Fitting a Gaussian to the spatial profile of the
star (along the slit), we derive $0.8$ arc-sec seeing (FWHM). Assuming
image symmetry, a 2-D calculation indicates $21\%$ loss at the slit
($0.43$ arc-sec slit width).  We thereby derive a line intensity of
$\sim 1.4 \times 10^{-16}$ ergs cm$^{-2}$ sec$^{-1}$.  Following the
calculation outlined by \citet{bary03} for optically thin emission,
this implies a mass $\sim 10^{-10}~M_{\bigodot}$ of hot H$_2$, and a
column density of $\sim 6 \times 10^{14}$ molecules cm$^{-2}$. This is
comparable to the hot H$_2$ mass in the near vicinity of several
weak-lined T-Tauri stars cited by \citet{bary03}, except that we argue
for shock-heating by outflow in KH~15D, not UV fluorescence or heating
by X-ray flux.  Note that the inferred mass is proportional to the
square of the source distance, and we have adopted $d=760$ pc.  Note
also (see below) that \citet{tok03} have imaged an extended jet of
S(1) emission, and the mass we derive here applies only to the
immediate vicinity of the star (within $\pm 0.9 \times 0.43$
arc-sec$^{2}$).

\section{Discussion}

\citet{tok03} have imaged KH~15D in a narrow band filter centered on
the 1-0 S(1) line.  They find a prominent jet starting in close
coincidence with the star, and extending many arc-sec to the north.
They have considered whether the jet is physically associated with
KH~15D or is a chance superposition (H$_2$ emission is common in
star-forming regions like NGC 2264).  Independent of our spectroscopy,
they conclude that the association of the jet with KH~15D is probably
physical.  We find that: 1) our low resolution spectrum during eclipse
shows bright S(1) emission centered on the stellar position to $\sim
0.2$ arc-sec (Sec. 3.1), and 2) our high resolution spectrum resolves
the line into two velocity components, whose relative intensities
change aburptly at the stellar position (Sec 3.2).  Therefore our
results strongly support the conclusion that S(1) emission is physically
associated with KH~15D.

Recently, quiescent H$_2$ 1-0~S(1) emission has been detected near
several T-Tauri stars \citep{wein00, bary03}. This quiescent emission
shows narrow line profiles, just as we see in KH~15D.  We considered
the possibility that the KH~15D emission originates from quiescent
gas, with Doppler shifts resulting from Keplerian orbital motion.  We
cannot exclude that some fraction of the emission, especially in the
red component, arises from quiescent emission.  Nevertheless, because
the KH~15D emission is spatially extended and shows evidence of
outflow, it seems more closely related to the shock-heated H$_2$ which
has been observed in hundreds of young stellar objects, including
classical T-Tauri stars \citep{bach96, her97}. 

Interpretation of our KH~15D observations requires knowing the stellar
radial velocity, and the heliocentric radial velocity we adopt for the
system ($+10$ km sec$^{-1}$) may be in error if the star is a binary.
Nevertheless, the double-peaked profiles we observe in the close
vicinity of the star suggest outflow.  However, a simple picture of a
symmetric bipolar jet emergent normal to a single edge-on disk is hard
to reconcile with our observations, and with the \citet{tok03}
imaging.  In such a simple interpretation, the line-of-sight shock
velocity in the 1-0 S(1) line would have to be at least half of the
difference between the red- and blueshifted components, i.e. $\gtrsim
30$ km sec$^{-1}$.  For a disk inclination of $\sim 84^{\circ}$
\citep{ham03}, the flow would be inclined by only $\sim 6^{\circ}$ to
the plane of the sky, so the total shock velocity would have to exceed
$\sim 250$ km sec$^{-1}$.  This is well above the velocity where
shocked H$_2$ is observed in other objects (H$_2$ becomes dissociated
at high shock velocities, \citet{sb82}).  Therefore the outflow
revealed in molecular hydrogen is likely to be inclined significantly to
the plane of the sky.  Similar arguments would not apply to lines
formed very close to the star, where a bipolar outflow may not be
collimated.  However, the emission we observe is spatially resolved
from the star, occurring at distances where bipolar outflows are often
highly collimated to a single axis.

\citet{tok03} find only one bright jet of line emission extending to
the north.  Since our slit was oriented primarily east-west (PA
$=66^{\circ}$), we have observed approximately (but not precisely)
perpendicular to the jet axis.  Our spectra will reveal the flow
pattern in the jet to the extent that the total angle between our slit
and the velocity vector differs from ${\pi}/2$.  We believe that our
brightest emission component (blueshifted `east' spectrum in Figure~2)
samples the \citet{tok03} jet, whereas the red components are
consistent with either a possible fainter counter-jet (not yet
detected by imaging), or quiescent emission near the stellar
velocity.  

The extent and complexity of H$_2$ emission from KH~15D is reminiscent
of T~Tauri itself, a known binary system.  Consistent with its binary
nature, T~Tauri shows two nearly perpendicular outflow systems
\citep{her97}.  Our observations are not yet sufficient, in terms of
spatial sampling, to say whether KH~15D shows a single, or multiple,
velocity systems.  Mapping spectroscopy of KH~15D in the 1-0~S(1) line
is needed to clarify the kinematics of this interesting object.

\section{Acknowledgements}

We thank Tim Brown for help with the observations, and the Keck
support staff, particularly Randy Campbell and Grant Hill, for
assistance with NIRSPEC.  We are grateful to Alan Tokunaga and
colleagues for communicating and discussing their imaging results, and
to an anonymous referee for several helpful comments. Portions of this
work were supported by NASA's Origins of Solar Systems program.  The
authors wish to recognize and acknowledge the very significant
cultural role and reverence that the summit of Mauna Kea has always
had within the indigenous Hawaiian community.  We are most fortunate
to have the opportunity to conduct observations from this mountain.

\clearpage

\begin{table}
\caption{Observed line intensities, normalized to 1-0 S(1), and
compared to the intensities expected from thermal excitation at
$T=2800K$}
\vspace{2mm}
\begin{tabular}{lccc}
\tableline
Line & Observed Intensity & Thermal Intensity ($T=2800K$) \\
\tableline
1-0 S(1)  & $1.0$ & $1.0$  \\
1-0 S(0)  & $0.30 \pm 0.04$ & $0.21$  \\
2-1 S(1) & $0.24 \pm 0.05$ & $0.19$ \\
1-0 Q(1) & $0.93 \pm 0.09$ & $0.71$  \\
1-0 Q(3) & $0.94 \pm 0.11$ & $0.80$   \\
\tableline
\tableline
\end{tabular}
\end{table}
\clearpage

\begin{figure}
\epsscale{0.7}
\plotone{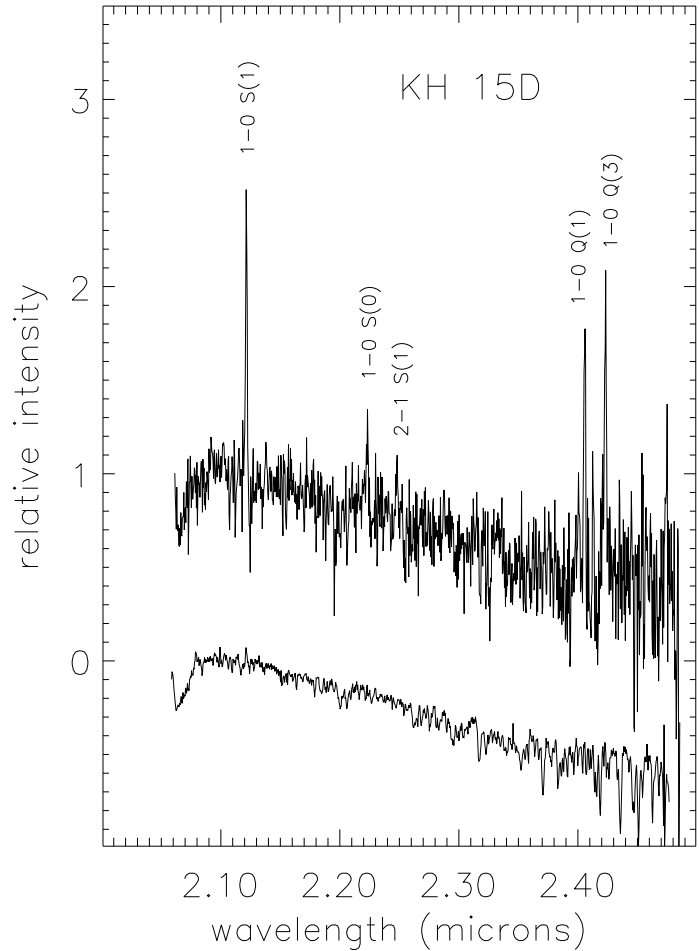}
\vspace{1.0 cm}
\caption{Keck/NIRSPEC low spectral resolution observations of KH~15D
during eclipse (upper, 19 August 2002) and out of eclipse (lower, 26
August 2002). The lower spectrum has been offset downward by 1.0 for
clarity.  Five emission lines due to molecular hydrogen are
identified, and are much more prominent relative to the stellar
continuum during eclipse.  The absorption lines which are most apparent in
the lower spectrum are telluric. \label{fig1}}
\end{figure}

\clearpage

\begin{figure}
\epsscale{0.7}
\plotone{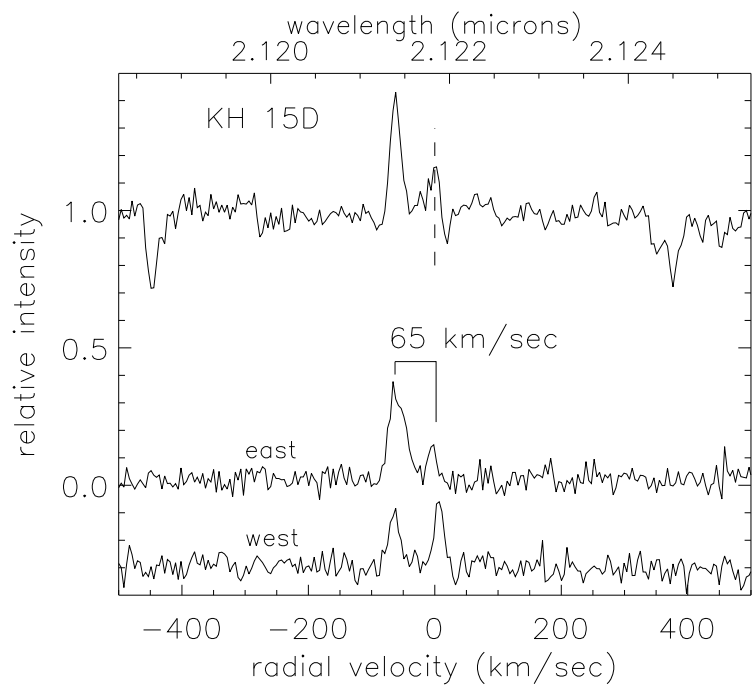}
\caption{Keck/NIRSPEC high spectral resolution observations of KH~15D
out of eclipse, showing double-peaked emission in the 1-0~S(1)
line. The upper spectrum was produced by integrating along the slit (oriented
approximately east-west) by $\pm 0.9$ arc-sec from the star, so it
shows the emission superposed on the stellar continuum.  The
absorption lines at $-450$ and $+370$ km sec$^{-1}$ are telluric.  The
two bottom spectra are offset east and west from the stellar position,
both produced by integrating along the slit $1.1-2.3$ arc-sec from the star.  (The
`west' spectrum has been offset downward by 0.3 for clarity.)  The
wavelength and radial velocity scales have been transformed into the
rest frame of the star, where the dashed vertical line marks zero
velocity.
\label{fig2}}
\end{figure}

\clearpage






\end{document}